\documentclass[12pt,sec]{article}
\usepackage{epsfig}
\usepackage{axodraw}
\begin{document}

\begin{center}
{\large  Inclusive Charmonium Production via Double $c \bar c$ in
$e^+e^-$  Annihilation}\\[0.8cm]
{ Kui-Yong Liu}

{\footnotesize Department of Physics, Peking University,
 Beijing 100871, People's Republic of China
 and Department of Physics, Liaoning University, Shenyang 110036, People's Republic of China}\\[0.5cm]

{ Zhi-Guo He}

{\footnotesize Department of Physics, Peking University,
 Beijing 100871, People's Republic of China}\\[0.5cm]

{ Kuang-Ta Chao}

{\footnotesize China Center of Advanced Science and Technology
(World Laboratory), Beijing 100080, People's Republic of China and
Department of Physics, Peking University,
 Beijing 100871, People's Republic of China}

\end{center}

\begin{abstract}

Motivated by the recent observation of double charm quark pair
production by the Belle Collaboration, we calculate the complete
${\cal O}(\alpha_{s}^{2})$ inclusive production cross sections for
$\eta_{c}$, $J/\psi$, and $\chi_{cJ}$(J=0, 1, 2) plus $c\bar{c}$
in $e^+ e^-$ annihilation through a virtual photon. We consider
both color-singlet and color-octet contributions, and give the
analytical expressions for these cross sections. The complete
color-singlet calculations are compared with the approximate
fragmentation calculations as functions of the center-of-mass
energy $\sqrt{s}$. We find that most of the fragmentation results
substantially overestimate the cross sections (e.g. by a factor of
$\sim$4 for $\chi_{c1}$ and $\chi_{c2}$) at the Belle and BaBar
energy $\sqrt{s}=10.6$GeV. The fragmentation results become a good
approximation only when $\sqrt{s}$ is higher than about $100$GeV.
We further calculate the color-octet contributions to these cross
sections with analytical expressions. We find that while the
color-octet contribution to $J/\psi$ inclusive production via
double charm is negligible (only about $3\%$), the color-octet
contributions to $\chi_{c1}$ and $\chi_{c2}$ can be significant.

PACS number(s): 12.40.Nn, 13.85.Ni, 14.40.Gx

\end{abstract}
\section{Introduction}
Charnonium is one of the simplest quark-antiquark composite
particles. Charmonium physics has played an important role in the
study of Quantum Chromodynamics (QCD) both perturbatively and
nonperturbatively, since the first charmonium state $J/\psi$ was
discovered in 1974. During the past decade, the study of
charmonium has become more interesting because of the large
difference between the predictions of the color-singlet model and
the observations of $J/\psi$ and $\psi'$ production at several
experimental facilities e.g. at the Fermilab Tevatron \cite{cdf}.

The newly developed nonrelativistic QCD (NRQCD) factorization
formalism \cite{bbl} allows the infrared safe calculation of
inclusive heavy quarkonium production and decay rates. In the
NRQCD production mechanism, a heavy quark-antiquark pair can be
produced at short distances in a conventional color-singlet or a
color-octet state, and then evolves into an observed quarkonium
nonperturbatively. With this color-octet mechanism, one may
explain the Tevatron data on the surplus production of $J/\psi $
and $\psi ^{\prime }$ at large $p_T$, though puzzles about their
polarizations still remain (for a recent review see \cite{kramer}
and references therein).

To further test the color octet mechanism, it is interesting to
study the charmonium production in $e^+e^-$ annihilation. The
$J/\psi$ inclusive production in $e^+e^-$ annihilation has been
investigated within the color-singlet model \cite{cm1,cm2,cm3} and
the color-octet model \cite{om1,om2,ko}. The angular distribution
and energy distribution of color-singlet $J/\psi$ production at
$\sqrt{s}=10.6~{\rm GeV}$ have been discussed in \cite{cm3}. In
\cite{om1} it is found that a clean signature of the color-octet
mechanism may be observed in the angular distribution of $J/\psi$
production near the end point region. In \cite{om2} contributions
of various color-octet as well as color-singlet channels to the
$J/\psi$ production cross sections are calculated in a wide range
of $e^+e^-$ collider energies. Moreover, the $J/\psi$
polarizations are predicted in \cite{ko}. Recently, BaBar
\cite{babar} and Belle \cite{belle} have measured the direct
$J/\psi$ production in continuum $e^+e^-$ annihilations at
$\sqrt{s}=10.6~{\rm GeV}$. The total cross section and the angular
distribution seem to favor the NRQCD calculation over the
color-singlet model \cite{babar}, but some issues (e.g. about the
momentum distribution and polarization of $J/\psi$) still remain.

The situation has become even more complicated due to the very
recent observation for the double $c\bar c$ production associated
with $J/\psi$ by Belle\cite{exdou}. The measured exclusive cross
section for $e^+ + e^-\rightarrow J/\psi+\eta_c$ process is an
order of magnitude larger than the theoretical value\cite{double},
and the measured inclusive cross section for $e^+ + e^-\rightarrow
J/\psi+c\bar c$ ($\sim$0.9 pb)\cite{belle,exdou} is more than
five times larger than NRQCD predictions which are only about
0.1-0.2 pb\cite{cm3,om2,ko,kiselev} taking into account the
differences in the values of the input parameters or methods.
Among other attempts to solve the $J/\psi c\bar c$ inclusive
production problem\cite{ioffe,berezhnoy},  $e^+ e^-$ annihilation
into two photons was also studied\cite{liu,luchinsky}, but the two
photon contribution turned to be negligible at $\sqrt{s}=10.6$GeV,
though it could prevail over one photon contribution at higher
energies (say, $\sqrt{s}>20$GeV)\cite{liu}.

The double $c\bar c$ production associated with $J/\psi$ (both
exclusively and inclusively) is very puzzling and needs a better
understanding for both perturbative and nonperturbative QCD. On
the other hand, experimentally, it is not clear whether the
copious (even dominant) double $c\bar c$ production will also
happen for charmonium states other than $J/\psi$, such as
$\eta_{c}$ and $\chi_{cJ}$(J=0, 1, 2). Among them the $\chi_{c1}$
and $\chi_{c2}$ are more interesting since they have large
branching fractions decaying into $J/\psi+\gamma$ and might be
easier to be detected. In fact, in Ref.~\cite{belle} the inclusive
production for $\chi_{c1}$ and $\chi_{c2}$ was searched for with
the available integrated luminosity of about 30 fb$^{-1}$ at
Belle. As more data are collected in the near future at B
factories we hope that more accurate measurements for the P-wave
and other S-wave charmonium states will be possible. These
measurements will be helpful to clarify the problems associated
with $J/\psi$ double $c\bar{c}$ production.

On the theoretical side, the calculations for inclusive S-wave and
P-wave charmonia production via double $c\bar{c}$ are necessary in
the framework of NRQCD, including both the color-singlet and
color-octet contributions. When we know the differences between
NRQCD predictions and experimental data, we will have to further
consider other mechanisms and methods in QCD to explain the
differences.

This paper is organized as follows. In Section 2, we will
calculate the complete $\cal{O}$$(\alpha_{s}^{2})$ color-singlet
inclusive production cross sections for $\eta_{c}$ and
$\chi_{cJ}$(J=0,1,2) (along with $J/\psi$) via double $c\bar{c}$
in $e^+ e^-$ annihilation through a virtual photon. Then we will
compare the complete calculation with the calculation obtained in
the charm quark fragmentation limit, and give their ratio as
functions of the center-of-mass energies and determine the energy
scales at which fragmentation approximations become reliable. In
section 3, we will further estimate the color-octet contributions
to $J/\psi$ and $\chi_{cJ}$ inclusive cross sections via double
charm. Finally, we summarize our results in section 4.

\section{Color-singlet contribution to charmonium production via double
$c\bar{c}$ in $e^+e^-$ annihilation}

The quarkonium can be described in term of Fock states
superposition within the NRQCD framework as follows

\begin{eqnarray}
&\mid \psi_{Q}>& =O(1)\mid Q\bar{Q}[^3S_1^{(1)}]> +
O(v)Q\bar{Q}[^3P_J^{(8)}]g> \nonumber\\&&+
O(v^2)Q\bar{Q}[^1S_0^{(8)}]g> + O(v^2)Q\bar{Q}[^3S_1^{(1,8)}]gg> +
\ldots,
\end{eqnarray}

\begin{eqnarray}
&\mid \chi_{cJ}>& =O(1)\mid Q\bar{Q}[^3P_J^{(1)}]> +
O(v)Q\bar{Q}[^3S_1^{(8)}]g> \nonumber\\&&+
O(v^2)Q\bar{Q}[^3P_J^{(1,8)}]gg> + \ldots,
\end{eqnarray}
where the superscript 1 or 8 labels the color configuration of the
Fock Components.

Following the nonrelativistic QCD (NRQCD) factorization formalism,
the scattering amplitude of the process
$e^-(p_1)+e^+(p_2)\rightarrow \gamma^* \rightarrow
c\bar{c}(^{2S+1}L_J^{(1,8a)})(p)+c(p_c)+\bar{c}(p_{\bar{c}})$ in
Fig.~\ref{feynman} and Fig.~\ref{fey2} is given by
\begin{eqnarray}
\label{amp2}   &&\hspace{-1cm}{\cal
A}(e^-(p_1)+e^+(p_2)\rightarrow
c\bar{c}(^{2S+1}L_{J}^{(1,8a)})(p)+c(p_c)+\bar{c}(p_{\bar{c}}))=\sqrt{C_{L}}
\sum\limits_{L_{z} S_{z} }\sum\limits_{s_1s_2 }\sum\limits_{jk}
\nonumber
\\ & \times&\langle s_1;s_2\mid
S S_{z}\rangle \langle L L_{z};S S_{z}\mid J J_{z}\rangle\langle
3j;\bar{3}k\mid 1,8a\rangle\nonumber\\
&\times&\left\{
\begin{array}{ll}
{\cal A}(e^-(p_1)+e^+(p_2)\rightarrow
 c_j(\frac{p}{2};s_1)+\bar{c}_k(\frac{p}{2};s_2)+
 c_l(\frac{p_c}{2};s_3)+\bar{c}_i(\frac{p_{\bar{c}}}{2};s_4))&(L=S),\nonumber\\
\epsilon^*_{\alpha}(L_Z) {\cal
A}^\alpha(e^-(p_1)+e^+(p_2)\rightarrow
 c_j(\frac{p}{2};s_1)+\bar{c}_k(\frac{p}{2};s_2)+
 c_l(\frac{p_c}{2};s_3)+\bar{c}_i(\frac{p_{\bar{c}}}{2};s_4))
&(L=P).
\end{array}
\right.\nonumber\\
\end{eqnarray}
where $c\bar{c}(^{2S+1}L_{J}^{(1,8a)})$ is the intermediate
$c\bar{c}$ pair produced at short distance, which subsequently
evolves into a specific charmonium state at long distance, ${\cal
A}^\alpha$ is the derivative of the amplitude with respect to the
relative momentum between the quark and anti-quark in the bound
state. For the case of color-singlet state, the coefficient
$C_{L}$ can be related to the origin of the radial wave function
(or its derivative) of the bound state as
\begin{equation}
C_{S}=\frac{1}{4\pi}\mid R_{S}(0) \mid^{2},~~~
C_{P}=\frac{3}{4\pi}\mid R_{P}'(0) \mid^{2}.
\end{equation}

The spin projection operator can be defined as\cite{pro}
\begin{equation}
P_{SS_z}(p;q)\equiv\sum\limits_{s_1s_2 }\langle
s_1;s_2|SS_z\rangle
v(\frac{p}{2}+q;s_1)\bar{u}(\frac{p}{2}-q;s_2).
\end{equation}

We list the spin projection operators and their derivatives with
respect to the relative momentum, which we will use in the
calculations, as
\begin{equation}
P_{00}(p,0)=\frac{1}{2\sqrt{2}}\gamma_{5}(\not{p}+2m_{c}),
\end{equation}
\begin{equation}
P_{1S_Z}(p,0)=\frac{1}{2\sqrt{2}}\not{\epsilon}(S_z)(\not{p}+2m_{c}),
\end{equation}
\begin{equation}
P_{1S_z}^{\alpha}(p,0)=\frac{1}{4\sqrt{2}m_c}
[\gamma^{\alpha}\not{\epsilon}^*(S_z)(\not{p}+2m_c)-
(\not{p}-2m_c)\not{\epsilon}(S_z)\gamma^{\alpha}].
\end{equation}

For P-wave states we need further relations to reduce the
polarizations
\begin{eqnarray}
\sum\limits_{L_ZS_Z}\epsilon^{*\alpha}(L_Z)\epsilon^{*\beta}(S_Z)\langle
1 L_Z;1 S_Z \mid J=0 J_Z=0
\rangle=\frac{1}{\sqrt{3}}(-g^{\alpha\beta}+\frac{p^{\alpha}p^{\beta}}{M^2})\nonumber\\
\sum\limits_{L_ZS_Z}\epsilon^{*\alpha}(L_Z)\epsilon^{*\beta}(S_Z)\langle
1 L_Z;1 S_Z \mid J=1 J_Z
\rangle=-\frac{i}{\sqrt{2}M}\epsilon^{\alpha\beta\lambda\kappa}p_{\kappa}\epsilon^{*}_{\lambda}(J_z)\\
\sum\limits_{L_ZS_Z}\epsilon^{*\alpha}(L_Z)\epsilon^{*\beta}(S_Z)\langle
1 L_Z;1 S_Z \mid J=1 J_Z
\rangle=\epsilon^{*\alpha\beta}(J_Z)\nonumber
\end{eqnarray}
where M is the mass of the charmonium, which equals to $2m_c$ in
the nonrelativistic approximation.

The calculation of cross sections for $e^{-}+e^{+}\rightarrow
\gamma^*\rightarrow$ charmonium $ + c\bar{c}$ is straightforward.
Using the definition in Ref.~\cite{cm3} we get the differential
cross section as follows
\begin{equation}
\label{cross} \frac{d\sigma(e^{+}+e^{-}\rightarrow \gamma^{*}
\rightarrow {\rm charmonium} +
c\bar{c})}{dz}=\frac{4C_{L}\alpha^{2}\alpha_{s}^{2}}{81m_{c}}(S(z)+\frac{\alpha(z)}{3}).
\end{equation}
where L=S for S-wave, L=P for P-wave and
$z={2E}_{J/\psi}/\sqrt{s}$. The functions $S(z)$ and $\alpha(z)$
for different charmonium states are given in the Appendix.

With Eq.~(\ref{cross}) we can evaluate the inclusive cross
sections for $\eta_{c}$, $J/\psi$ and $\chi_{cJ}$. The input
parameters used in the numerical calculations are\cite{wf}
\begin{equation}
m_e=0, ~~m_c=1.5~{\rm GeV}, ~~\alpha_s(2m_c)=0.26,~~\alpha=1/137,
\end{equation}
\begin{equation}
\mid R_S(0) \mid^2=0.81~{\rm GeV}^3,~~~\mid R_P(0)'
\mid^2=0.075~{\rm GeV}^5.
\end{equation}

Now we give the numerical results at the Belle and BaBar energy
$\sqrt{s}=10.6$ GeV.

\begin{equation}
\label{etc}
\sigma(e^{+}+e^{-}\rightarrow\gamma^{*}\rightarrow\eta_{c}+c\bar{c})=58.7~
{\rm fb}
\end{equation}

\begin{equation}
\label{js} \sigma(e^{+}+e^{-}\rightarrow\gamma^{*}\rightarrow
J/\psi+c\bar{c})=148~{\rm fb}
\end{equation}

\begin{equation}
\label{kc0}
\sigma(e^{+}+e^{-}\rightarrow\gamma^{*}\rightarrow\chi_{c0}+c\bar{c})=48.8~
{\rm fb}
\end{equation}

\begin{equation}
\label{kc1}
\sigma(e^{+}+e^{-}\rightarrow\gamma^{*}\rightarrow\chi_{c1}+c\bar{c})=13.5~
{\rm fb}
\end{equation}

\begin{equation}
\label{kc2}
\sigma(e^{+}+e^{-}\rightarrow\gamma^{*}\rightarrow\chi_{c2}+c\bar{c})=6.30~
{\rm fb}
\end{equation}

The $J/\psi$ production rate is in agreement with other
references\cite{cm3,om2,ko} after taking into account the
differences in the values of the input parameters. In the
$z\gg\delta$ limit, where $\delta$ is defined as $4m_c/\sqrt{s}$,
the approximate fragmentation results will be equivalent to the
complete calculations. This is another check for the validity of
the complete calculation. Here the fragmentation cross sections
are written as

\begin{eqnarray}
\label{frag}
\sigma_{frag}(e^{+}+e^{-}\rightarrow\gamma^{*}\rightarrow{
charmonium}+c\bar{c})=\nonumber\\2\sigma(e^{+}+e^{-}\rightarrow
c\bar{c})\int_{\delta}^{1}{\cal D}_{c\rightarrow{
charmonium}}(z)dz,
\end{eqnarray}
where  ${\cal D}(z)$ are the charm quark fragmentation functions
into S-wave\cite{s} or P-wave\cite{p} charmonia.

The cross sections obtained in the complete calculation and in the
fragmentation approximation as functions of the center-of-mass
energies are plotted in Fig.~\ref{fig1}-\ref{fig5}. All these
cross sections are in units of $\sigma_{cc}=
\sigma(e^{+}+e^{-}\rightarrow\gamma^{*}\rightarrow c\bar{c})$,
 the cross section for $e^{+}e^{-}$
annihilating into the $c\bar{c}$ quark pair, times $10^{-4}$. One
can find that the cross sections in complete calculations and
fragmentation approximations (all in units of the cross section
for $e^{+}e^{-}$ annihilating to the $c\bar{c}$ pair) are
proportional to the fragmentation probabilities for the charm
quark fragmentating into charmonia when the $\delta\ll 1$ limit is
valid. This is just what the fragmentation approach describes. The
results in these figures show that except for $\chi_{c0}$, the
differences between fragmentation results and complete
calculations are large at low energies. At the Belle and BaBar
energy $\sqrt{s}$=10.6GeV, the ratios of complete calculations to
fragmentation calculations are
\begin{equation}
\frac{\sigma(e^{+}+e^{-}\rightarrow\gamma^{*}\rightarrow
charmonium+c\bar{c})}{\sigma_{frag}(e^{+}+e^{-}\rightarrow\gamma^{*}\rightarrow
charmonium+c\bar{c})}=0.28, 0.58, 0.25, 0.25
\end{equation}
for $\eta_{c}$, $J/\psi$, $\chi_{c1}$, and $\chi_{c2}$
respectively. As the center-of-mass energy increases, the ratios
of complete calculations to fragmentation results increase and can
reach to $90\%$ when the center-of-mass energy is over 100 GeV.
Moreover, the cross sections are rather sensitive to the input
parameters. If we choose $\alpha=1/134, \alpha_{s}=0.28$, and
${\rm m_{c}}=1.48{\rm GeV}$ at $\sqrt{s}=10.6$GeV, the cross
sections for $\eta_{c}, J/\psi$, and $\chi_{cJ}$ (J=0,1,2) become
77.0fb, 192fb, 64.2fb, 18.3fb, 8.48fb respectively.

\section{Color-octet contribution to $J/\psi$ and $\chi_{cJ}$
production via double $c\bar{c}$ in $e^+e^-$ annihilation}

\begin{figure}
\begin{center}
\includegraphics[width=14cm,height=18cm]{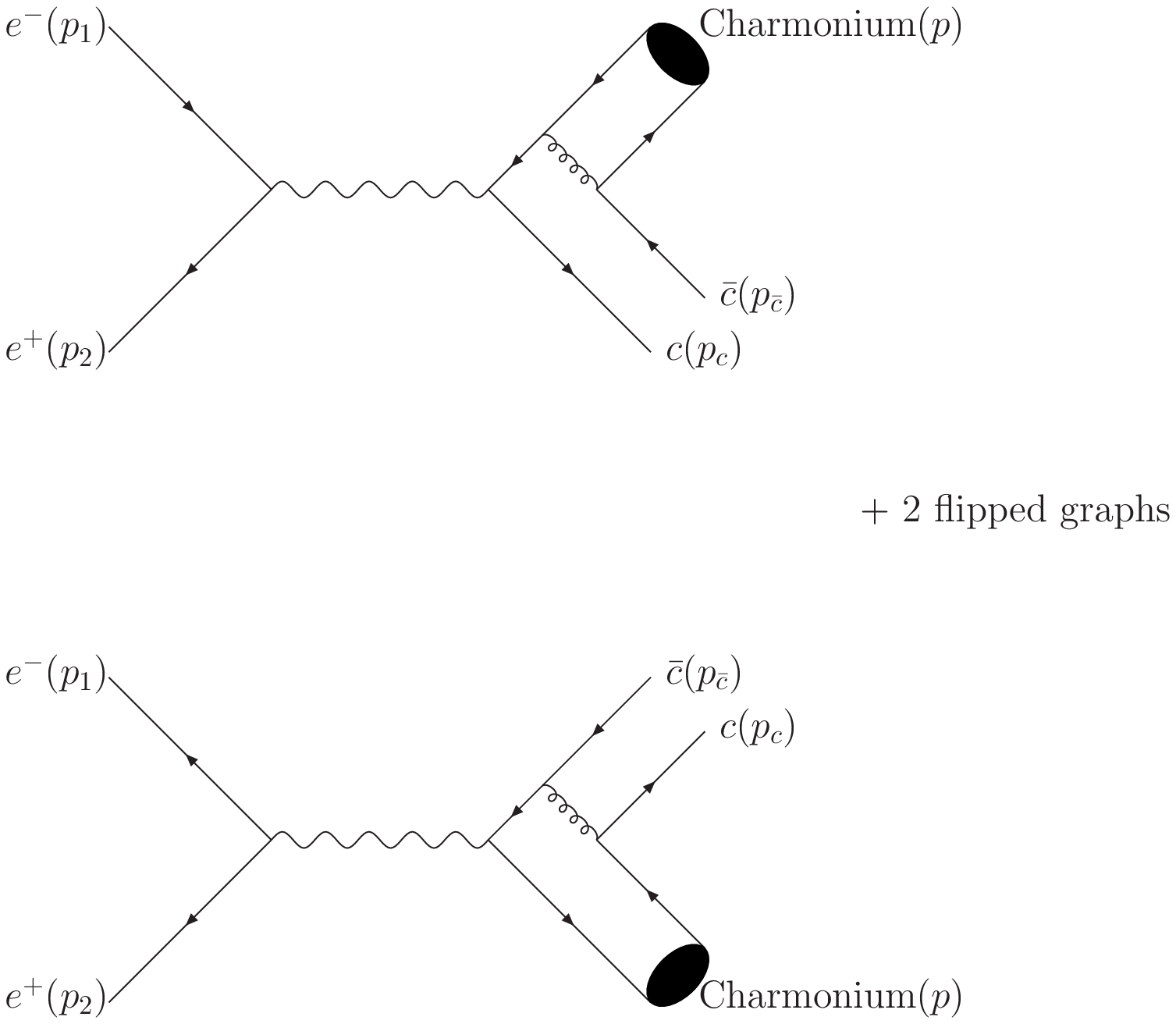}
\end{center}
\caption{ Feynman diagrams for $e^+ +
e^-\rightarrow\gamma^*\rightarrow$ Charmonium + $c\bar{c}$.}
\label{feynman}
\end{figure}

\begin{figure}
\begin{center}
\vspace{-3.0cm}
\includegraphics[width=14cm,height=18cm]{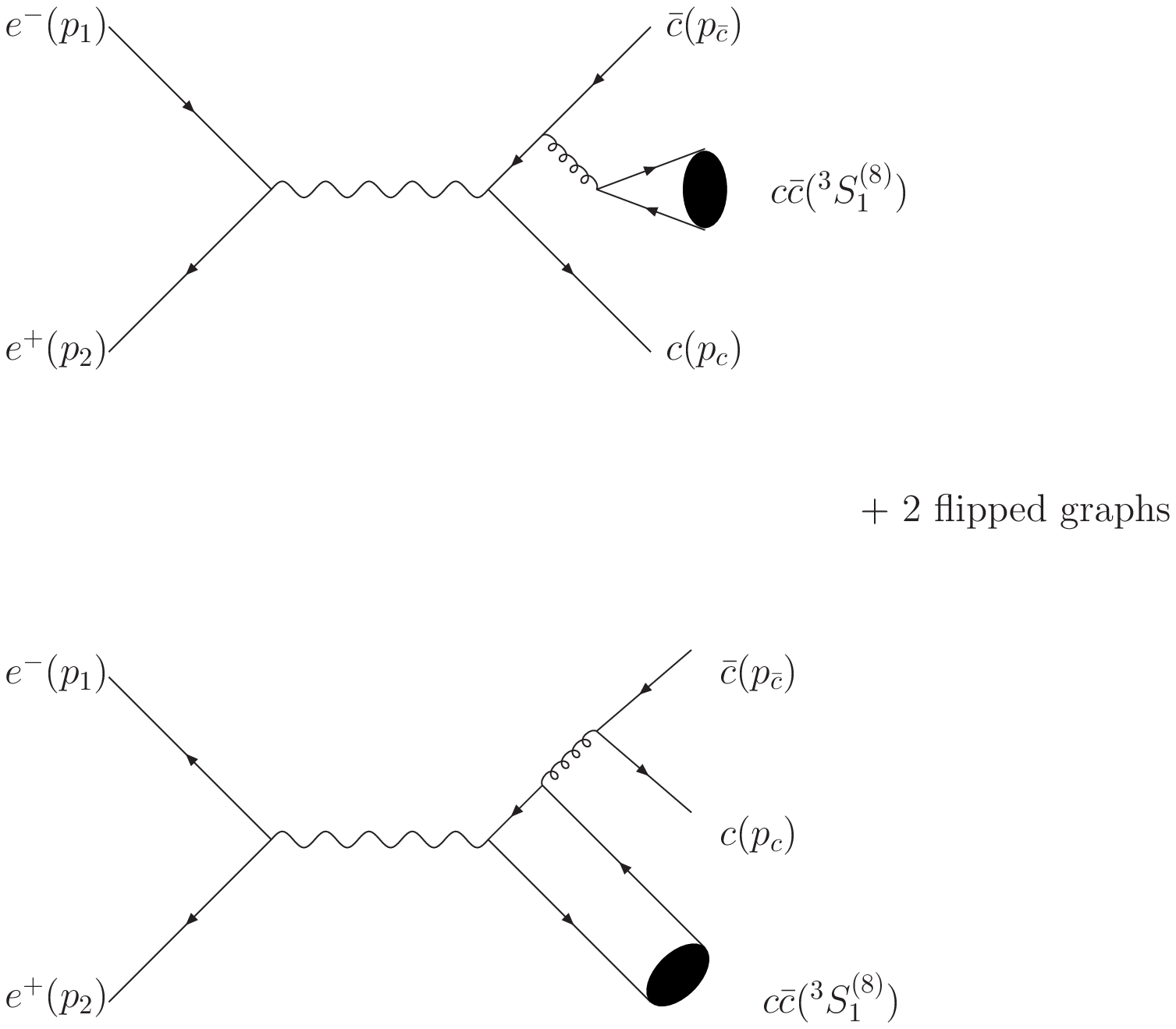}
\vspace{-4cm}
\end{center}
\caption{ Feynman diagrams for $e^+ +
e^-\rightarrow\gamma^*\rightarrow$ $c\bar{c}(^{2S+1}L_J^{(8)})$ +
$c\bar{c}$.} \label{fey2}
\end{figure}

\begin{figure}
\begin{center}
\vspace{4.0cm}
\includegraphics[width=14cm,height=8cm]{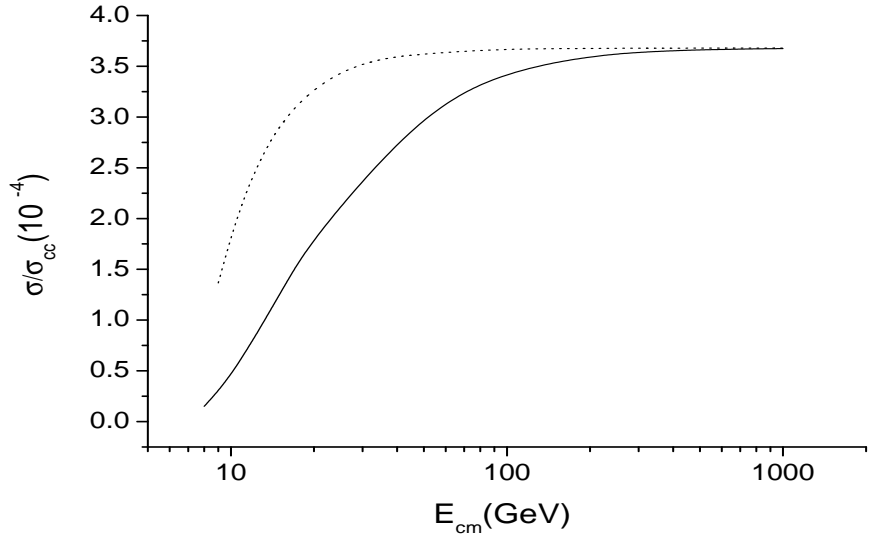}
\vspace{-0cm}
\end{center}
\caption{ Cross sections for $e^+ e^-\rightarrow \eta_{c} +
c\bar{c}$ plotted against the center-of-mass energy. Dotted line
illustrates the fragmentation calculation and solid line
illustrates the complete calculation. The cross sections are in
units of $\sigma_{cc}=
\sigma(e^{+}+e^{-}\rightarrow\gamma^{*}\rightarrow c\bar{c})$
times $10^{-4}$.} \label{fig1}
\end{figure}

\begin{figure}
\begin{center}
\vspace{5.0cm}
\includegraphics[width=14cm,height=8cm]{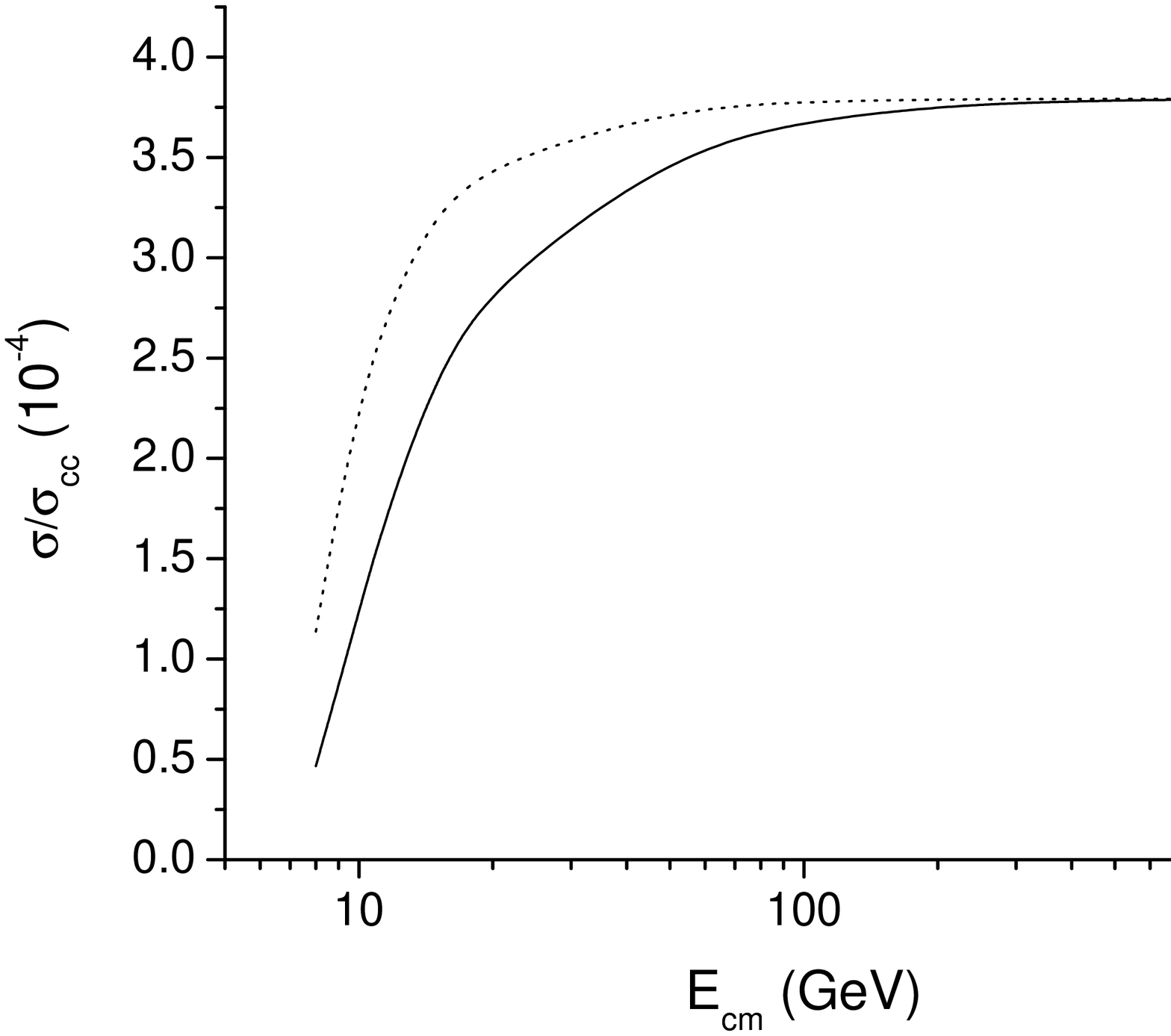}
\vspace{-0cm}
\end{center}
\caption{ Cross sections for $e^+ e^-\rightarrow J/\psi +
c\bar{c}$ plotted against the center-of-mass energy. Dotted line
illustrates the fragmentation calculation and solid line
illustrates the complete calculation.} \label{fig2}
\end{figure}

\begin{figure}
\begin{center}
\vspace{5.0cm}
\includegraphics[width=14cm,height=8cm]{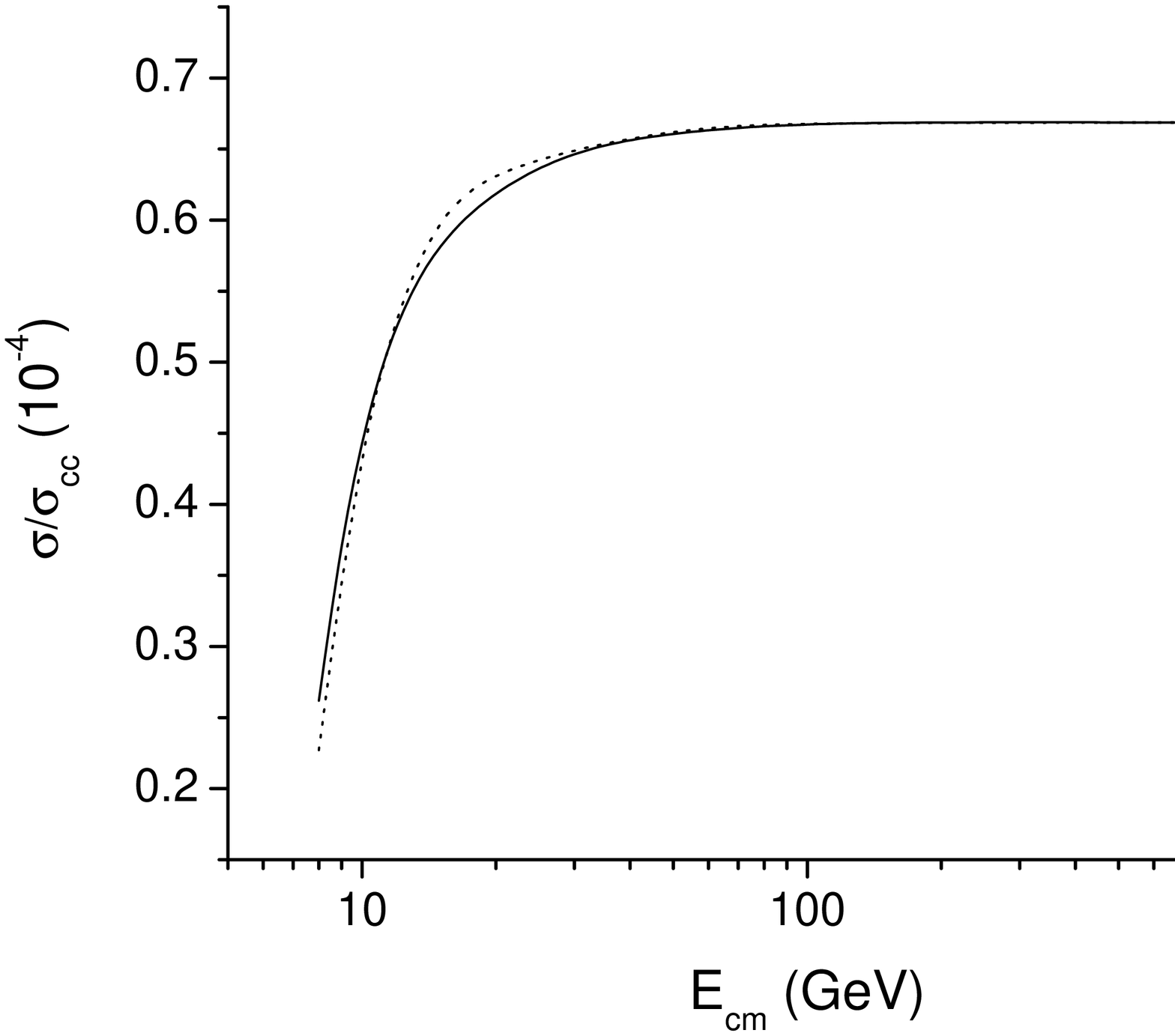}
\vspace{-0cm}
\end{center}
\caption{ Cross sections for $e^+ e^-\rightarrow \chi_{c0} +
c\bar{c}$ plotted against the center-of-mass energy. Dotted line
illustrates the fragmentation calculation and solid line
illustrates the complete calculation.} \label{fig3}
\end{figure}

\begin{figure}
\begin{center}
\vspace{5.0cm}
\includegraphics[width=14cm,height=8cm]{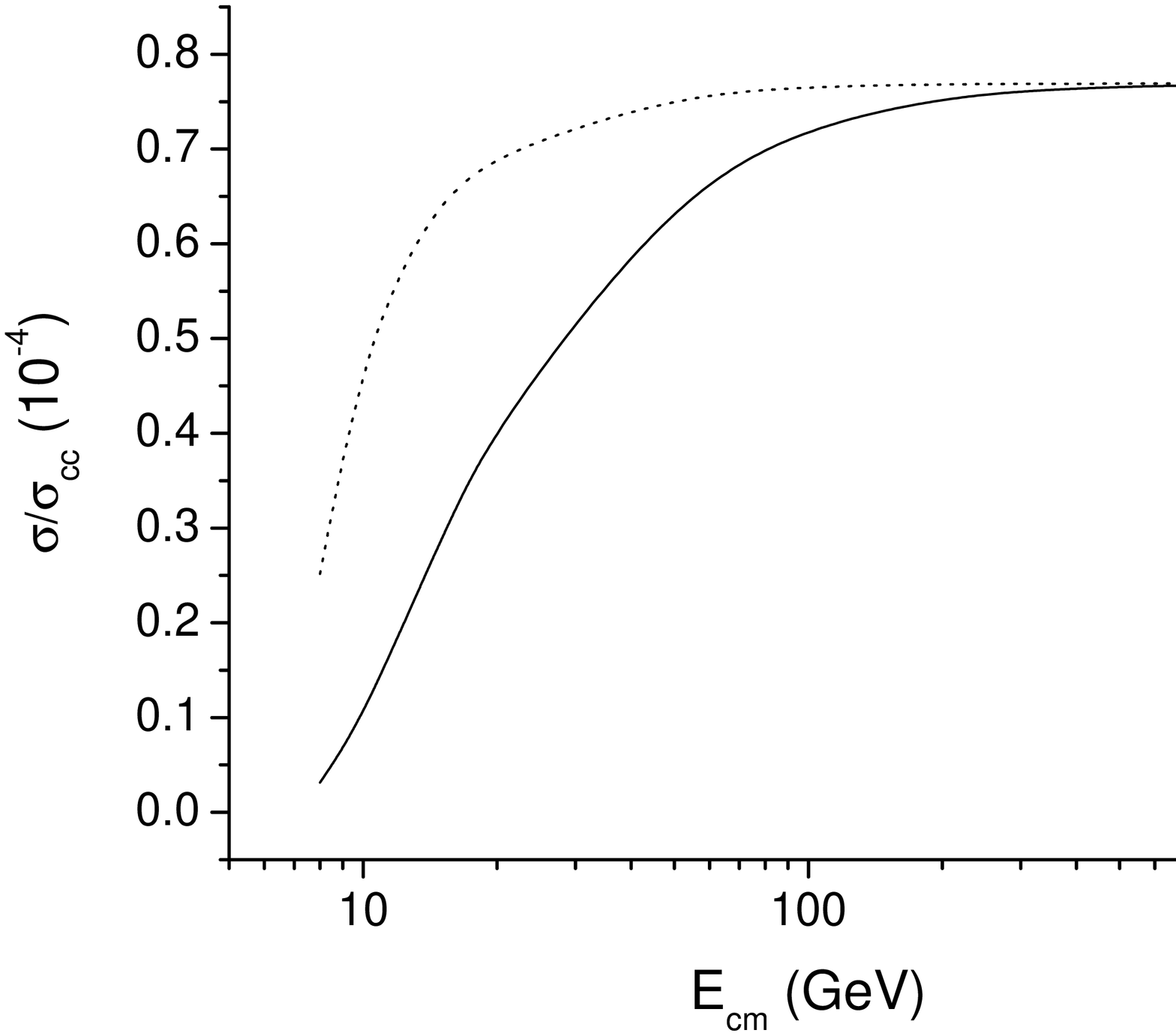}
\vspace{-0cm}
\end{center}
\caption{ Cross sections for $e^+ e^-\rightarrow \chi_{c1} +
c\bar{c}$ plotted against the center-of-mass energy. Dotted line
illustrates the fragmentation calculation and solid line
illustrates the complete calculation.} \label{fig4}
\end{figure}

\begin{figure}
\begin{center}
\vspace{5.0cm}
\includegraphics[width=14cm,height=8cm]{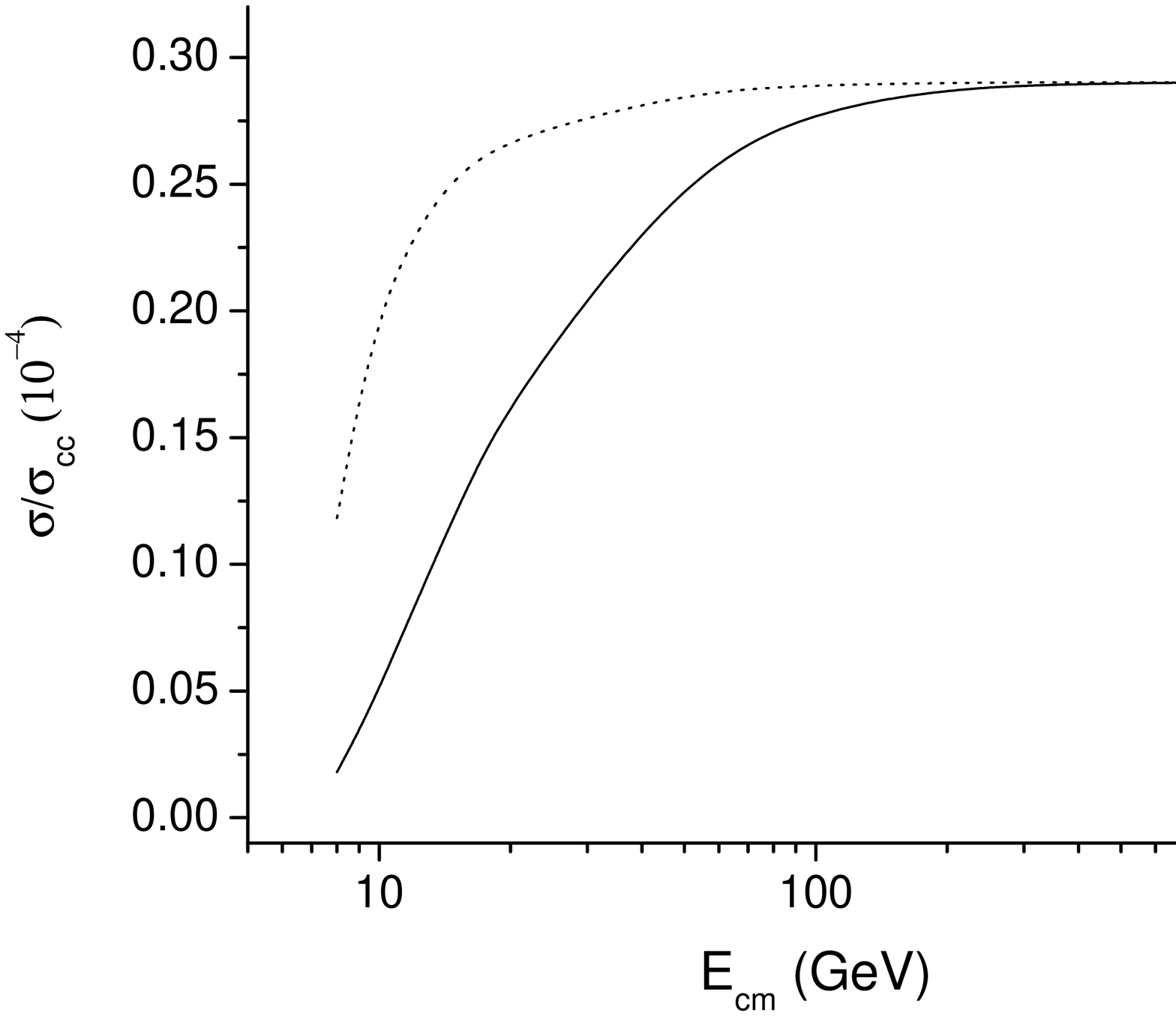}
\vspace{-0cm}
\end{center}
\caption{ Cross sections for $e^+ e^-\rightarrow \chi_{c2} +
c\bar{c}$ plotted against the center-of-mass energy. Dotted line
illustrates the fragmentation calculation and solid line
illustrates the complete calculation.} \label{fig5}
\end{figure}

We next estimate the color-octet contribution to $J/\psi$ and
$\chi_{cJ}$ production via double $c\bar{c}$ in $e^+e^-$
annihilation. The Feynman diagrams are showed in Fig~\ref{feynman}
and Fig~\ref{fey2}.

In Fig.~\ref{feynman}, the charmonium comes from the color-octet
mediate states $c\bar{c}(^{2S+1}L_J^{(8)})$ by emitting  soft
gluons at long distances. Here the color-octet contribution can be
obtained from the corresponding color-singlet contribution divided
by a factor of $\frac{32<{\cal O}^{H}_1(^{2S+1}L_{J})>}{3<{\cal
O}^{H}_8(^{2S+1}L_J)>}$. The matrix elements $<{\cal
O}^{H}_n(^{2S+1}L_J)>$ can be extracted from the Tevatron data for
$J/\psi$ and $\chi_{cJ}$ production (see Ref.~\cite{omes} for
detailed discussions). Accordingly, with some unavoidable
uncertainties we set them as

\begin{equation}
<{\cal O}^{\psi}_1(^{3}S_1)>=1.16~{\rm GeV}^3,
\end{equation}

\begin{equation}
<{\cal O}^{\chi_{c1}}_1(^{3}P_1)>=0.32~{\rm GeV}^5,
\end{equation}

\begin{equation}
<{\cal O}^{\psi}_8(^{3}S_1)>=1.06\times 10^{-2}~{\rm GeV}^3,
\end{equation}

\begin{equation}
<{\cal O}^{\psi}_8(^{3}P_0)>/m_c^2=1.0\times 10^{-2}~{\rm GeV}^3,
\end{equation}

\begin{equation}
<{\cal O}^{H}_8(^{3}P_J)>=(2J+1)<{\cal O}^{H}_8(^{3}P_0)>,
\end{equation}

\begin{equation}
<{\cal O}^{\chi_{c1}}_8(^{3}S_1)>=1.0\times 10^{-2}~{\rm GeV}^3.
\end{equation}

With these values of the matrix elements, the color-octet
contributions to $J/\psi$ and $\chi_{cJ}$ in Fig.~\ref{feynman}
are about at least two orders of magnitude smaller than the
color-singlet contributions, and therefore are negligible. In
Fig.~\ref{fey2} the color-octet contributions come from four
different (the upper two and the lower two) diagrams. With their
contributions (including the interference terms), the differential
cross section reads

\begin{equation}
\frac{d\sigma_{octet}}{dz}=\frac{16\alpha^2\alpha_s^2<{\cal
O}^{H}_8(^{3}S_1)>}{27m_c}\mid \bar{M} \mid^2,
\end{equation}
where $\mid \bar{M} \mid^2$ takes the form
\begin{eqnarray}
\mid \bar{M} \mid^2&=& \frac{\pi}{12 \delta^2 s^2 z
(z-2)^2}\{-4z\sqrt{\frac{(1-z)(z^2-\delta^2)}{4+\delta^2-4z}}\nonumber\\
&&[3\delta^4-12\delta^2(z-2)+16(10+z(z-10))]+
\nonumber\\
&&(z-2)^2[3\delta^4-8\delta^2(3z-4)+32(2+z(z-2))]\nonumber\\
&&{\rm ln}[\frac{z \sqrt{4 + \delta^2 - 4
z}+2\sqrt{(1-z)(z^2-\delta^2)}}{z \sqrt{4 + \delta^2 - 4
z}-2\sqrt{(1-z)(z^2-\delta^2)}}]\}.
\end{eqnarray}
The numerical results can be obtained by using the parameters
given above, and are

\begin{equation}
\sigma_{octet}(e^+e^-\rightarrow J/\psi c\bar{c})=4.5~{\rm fb}.
\end{equation}

\begin{equation}
\sigma_{octet}(e^+e^-\rightarrow \chi_{c1} c\bar{c})=4.3~{\rm fb}.
\end{equation}

The color-octet contribution to $J/\psi$ is only $3\%$ of the
color-singlet cross section. For $\chi_{c1}$, the color-octet
contribution is significant, which is about $32\%$ of the
color-singlet cross section. With the approximation of heavy quark
spin symmetry, the contributions of color octet  $ ^{3}S_1$ to
$\chi_{cJ}$ (from color-octet $^{3}S_1$ mediate state to
color-singlet $^{3}P_J$ final state by E1 transition) satisfy the
ratio $1:3:5$ for $J=0,1,2$ respectively. Their values are given
by

\begin{equation}
\sigma_{octet}(e^+e^-\rightarrow \chi_{c0} c\bar{c})=1.4~{\rm fb}.
\end{equation}

\begin{equation}
\sigma_{octet}(e^+e^-\rightarrow \chi_{c2} c\bar{c})=7.2~{\rm fb}.
\end{equation}

\begin{figure}
\begin{center}
\vspace{5.0cm}
\includegraphics[width=14cm,height=10cm]{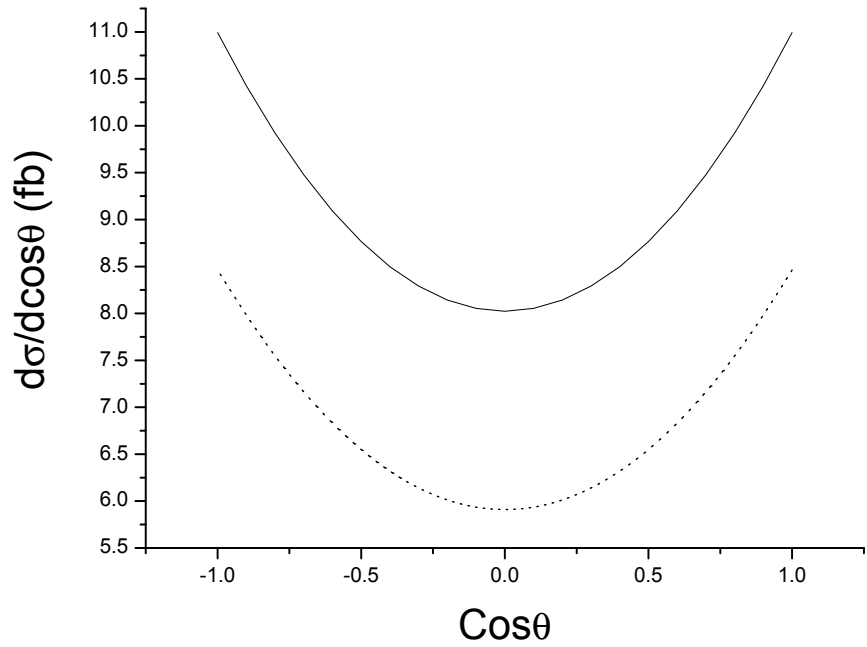}
\vspace{-0cm}
\end{center}
\caption{ Differential cross sections of the color-singlet (dotted
line) and the sum of color-singlet and color-octet (solid line)
contributions as functions of the production angle of
$\chi_{c1}$.} \label{ang}
\end{figure}

\begin{figure}
\begin{center}
\vspace{5.0cm}
\includegraphics[width=14cm,height=8cm]{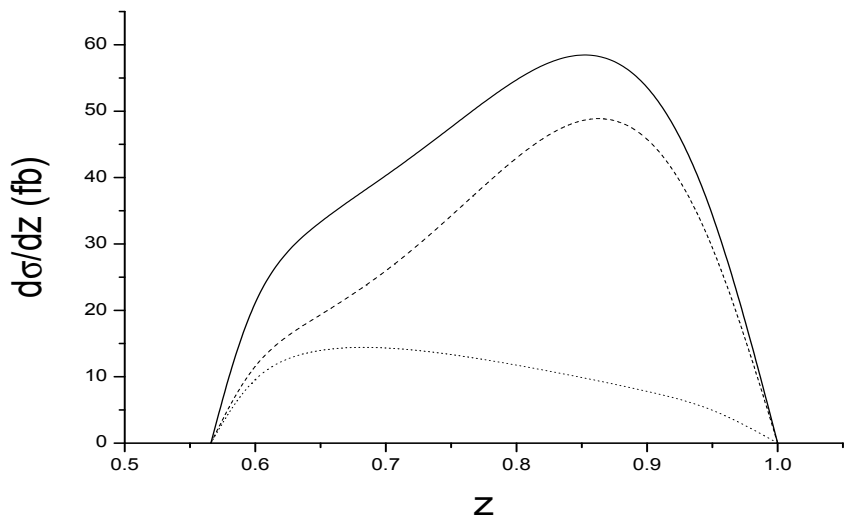}
\vspace{-0cm}
\end{center}
\caption{ Contributions to $d\sigma(e^+e^- \rightarrow
\chi_{c1}c\bar{c})/dz$ from color-singlet (dashed line),
color-octet (dotted line) and the sum of color-singlet and
color-octet (solid line) contributions plotted against $z$.}
\label{z3}
\end{figure}

We show the angular distribution and energy distribution for
$\chi_{c1}$ in Fig~\ref{ang} and Fig~\ref{z3}. One can see that
the color-octet contribution enhances the differential cross
section significantly in the low energy (small $z$) region.

\section{Conclusion}

In summary, we have calculated the complete ${\cal
O}(\alpha_{s}^{2})$ inclusive production cross sections for
$\eta_{c}$, $J/\psi$, and $\chi_{cJ}$(J=0, 1, 2) plus $c\bar{c}$
in $e^+ e^-$ annihilation through a virtual photon. We consider
both color-singlet and color-octet contributions, and give the
analytical expressions for these cross sections. The complete
color-singlet calculations are compared with the approximate
fragmentation calculations as functions of the center-of-mass
energy $\sqrt{s}$. We find that most of the fragmentation results
substantially overestimate the cross sections (e.g. by a factor of
$\sim$4 for $\chi_{c1}$ and $\chi_{c2}$) at the Belle and BaBar
energy $\sqrt{s}=10.6$GeV. The fragmentation results become a good
approximation only when $\sqrt{s}$ is higher than about $100$GeV.
We further calculated the color-octet contributions to these cross
sections with analytical expressions. We find that while the
color-octet contribution to $J/\psi$ inclusive production via
double charm is negligible (only about $3\%$), the color-octet
contributions to $\chi_{c1}$ and $\chi_{c2}$ can be significant.
These results may serve as NRQCD predictions to compare with the
experimental data observed or to be observed at Belle and BaBar.

\section*{Acknowledgments}

The authors thank Z.Z. Song for useful discussions. This work was
supported in part by the National Natural Science Foundation of
China, and the Education Ministry of China.

\section*{Appendix}

In this Appendix, we give the functions of S and $\alpha$ which
are defined in Eq.~(\ref{cross}).
\begin{eqnarray}
{\rm
S}_{\eta_c}&=&\frac{4\pi}{3s^{2}\delta^{2}z^{3}(z-2)^{6}(z^{2}-\delta^{2})}
\{4z\sqrt{\frac{(1-z)(z^{2}-\delta^{2})}{4+\delta^{2}-4z}}[-96\delta^{6}(2+\delta^{2})(4+\delta^2) \nonumber \\
&&
+96\delta^{6}(64+22\delta^{2}+\delta^{4})z-16\delta^{2}(1920-864\delta^{2}+532\delta^{4}+125\delta^{6}-2\delta^{8})z^{2}
\nonumber \\
&&+8\delta^{2}(9984-5312\delta^{2}+488\delta^{4}+96\delta^{6}-\delta^{8})z^{3} \nonumber \\
&&
+2(6144-47872\delta^{2}+20800\delta^{4}-392\delta^{6}-110\delta^{8}+3\delta^{10})z^{4} \nonumber \\
&&
-4(6144-21376\delta^{2}+4256\delta^{4}+112\delta^{6}+9\delta^{8})z^{5} \nonumber \\
&&
+(14336-51328\delta^{2}+5472\delta^{4}+420\delta^{6}-3\delta^{8})z^{6}\nonumber\\
&&-4(1536-3168\delta^{2}+352\delta^{4}+\delta^{6})z^{7} \nonumber\\
&&
+8(864-36\delta^{2}+13\delta^{4})z^{8}-32(112+11\delta^{2})z^{9}+768z^{10}] \nonumber \\
&&
-3\delta^{2}(z-2)^{4}[8\delta^{6}(2+\delta^{2})-96\delta^{6}z-2\delta^{2}(192-48\delta^{2}+8\delta^{4}-\delta^{6})z^{2}
\nonumber \\
&&
+16\delta^{2}(8+6\delta^{2}-\delta^{4})z^{3}+\delta^{2}(192+40\delta^{2}-\delta^{4})z^{4}
+8(32-4\delta^{2}+\delta^{4})z^{5} \nonumber\\
&&
-8(48+\delta^{2})z^{6}]\ln\frac{z\sqrt{4+\delta^{2}-4z}+2\sqrt{(1-z)(z^{2}-\delta^{2})}}
{z\sqrt{4+\delta^{2}-4z}-2\sqrt{(1-z)(z^{2}-\delta^{2})}}\}.
\end{eqnarray}

\begin{eqnarray}
{\rm
\alpha}_{\eta_c}&=&\frac{4\pi}{3s^{2}\delta^{2}z^{3}(z-2)^{6}(z^{2}-\delta^{2})}
\{4z\sqrt{\frac{(1-z)(z^{2}-\delta^{2})}{4+\delta^{2}-4z}}[96\delta^{6}(4+\delta^{2})(6+\delta^{2}) \nonumber \\
&& -96\delta^{6}(64+18\delta^{2}+\delta^{4})z+16\delta^{2}(2688+608\delta^{2}+428\delta^{4}
+43\delta^{6}-2\delta^{8})z^{2} \nonumber \\
&&
-8\delta^{2}(17664+3264\delta^{2}+184\delta^{4}-96\delta^{6}-\delta^{8})z^{3}
\nonumber \\
&&
+2(6144+89344\delta^{2}+7744\delta^{4}-2024\delta^{6}-174\delta^{8}-3\delta^{10})z^{4}
\nonumber \\
&&
-4(6144+22656\delta^{2}-1376\delta^{4}-512\delta^{6}-35\delta^{8})z^{5}
\nonumber \\
&&
+(14336+5504\delta^{2}-5152\delta^{4}-732\delta^{6}-3\delta^{8})z^{6}\nonumber\\
&&-4(1536-1760\delta^{2}-416\delta^{4}+\delta^{6})z^{7}
\nonumber \\
&&
+8(864-196\delta^{2}+13\delta^{4})z^{8}-32(112+11\delta^{2})z^{9}+768z^{10}]
\nonumber \\
&& +3\delta^{2}(z-2)^{4}[8\delta^{6}(6+\delta^{2})-32\delta^{6}z
-2\delta^{2}(64+48\delta^{2}+16\delta^{4}-\delta^{6})z^{2} \nonumber \\
&&
+16\delta^{2}(12-\delta^{2})(2+\delta^{2})z^{3}-(1024+320\delta^{2}-88\delta^{4}-\delta^{4})z^{4}\nonumber \\
&&+8(96-28\delta^{2}-\delta^{4})z^{5}+8(16+\delta^{2})z^{6}]\nonumber\\
&&\times\ln\frac{z\sqrt{4+\delta^{2}-4z}+2\sqrt{(1-z)(z^{2}-\delta^{2})}}
{z\sqrt{4+\delta^{2}-4z}-2\sqrt{(1-z)(z^{2}-\delta^{2})}}\}.
\end{eqnarray}

\begin{eqnarray}
{\rm
S}_{\psi}&=&\frac{4\pi}{s^{2}\delta^{2}z^{3}(z-2)^{6}(z^{2}-\delta^{2})}
\{4z\sqrt{\frac{(1-z)(z^{2}-\delta^{2})}{4+\delta^{2}-4z}}\nonumber\\
&&\times[-32\delta^{4}(4+\delta^{2})(48+22\delta^{2}+3\delta^{4}) \nonumber \\
&&
+32\delta^{4}(768+400\delta^{2}+66\delta^{4}+3\delta^{6})z\nonumber\\
&&-16\delta^{2}(384+1920\delta^{2}+556\delta^{4}+29\delta^{6}-2\delta^{8})z^{2}
\nonumber \\
&&
+8\delta^{2}(1792+128\delta^{2}-568\delta^{4}-80\delta^{6}-\delta^{8})z^{3} \nonumber \\
&&
+2(2048-11008\delta^{2}+10752\delta^{4}+3176\delta^{6}+98\delta^{8}+3\delta^{10})z^{4}
\nonumber \\
&&
-4(4096-7808\delta^{2}+3424\delta^{4}+600\delta^{6}+17\delta^{8})z^{5}
\nonumber \\
&&
+(38912-20608\delta^{2}+4544\delta^{4}+508\delta^{6}-3\delta^{8})z^{6}
\nonumber \\
&&
-4(13312-800\delta^{2}+120\delta^{4}-3\delta^{6})z^{7}+8(4512-20\delta^{2}-15\delta^{4})z^{8}
\nonumber \\
&& -32(336-\delta^{2})z^{9}+1280z^{10}]
\nonumber \\
&& -\delta^{2}(z-2)^{4}[8\delta^{4}(48+22\delta^{2}+3\delta^{4})-32\delta^{4}(24+5\delta^{2})z \nonumber \\
&& -2\delta^{2}(448+16\delta^{2}+8\delta^{4}-3\delta^{6})z^{2}+16\delta^{2}(56-10\delta^{2}-5\delta^{4})z^{3} \nonumber \\
&& +\delta^{2}(1152+272\delta^{2}-3\delta^{4})z^{4}+8(32-92\delta^{2}+5\delta^{4})z^{5}-56(16+\delta^{2})z^{6} \nonumber \\
&&
+512z^{7}]\ln\frac{z\sqrt{4+\delta^{2}-4z}+2\sqrt{(1-z)(z^{2}-\delta^{2})}}
{z\sqrt{4+\delta^{2}-4z}-2\sqrt{(1-z)(z^{2}-\delta^{2})}}\}.
\end{eqnarray}

\begin{eqnarray}
{\rm
\alpha}_{\psi}&=&\frac{4\pi}{s^{2}\delta^{2}z^{3}(z-2)^{6}(z^{2}-\delta^{2})}
\{4z\sqrt{\frac{(1-z)(z^{2}-\delta^{2})}{4+\delta^{2}-4z}}\nonumber\\
&&\times[32\delta^{4}(4+\delta^{2})(16+2\delta^{2}+3\delta^{4})
-32\delta^{4}(256+48\delta^{2}+22\delta^{4}+3\delta^{6})z \nonumber \\
&& +16\delta^{2}(1152+1024\delta^{2}-140\delta^{4}
-53\delta^{6}-2\delta^{8})z^{2} \nonumber \\
&&
-8\delta^{2}(5376+128\delta^{2}-1576\delta^{4}-240\delta^{6}-\delta^{8})z^{3}
\nonumber \\
&&
+2(2048-768\delta^{2}-19968\delta^{4}-6968\delta^{6}-350\delta^{8}-3\delta^{10})z^{4}
\nonumber \\
&&
-4(4096-20096\delta^{2}-11168\delta^{4}-1208\delta^{6}-43\delta^{8})z^{5} \nonumber \\
&&+(38912-75392\delta^{2}-16960\delta^{4}
-996\delta^{6}-3\delta^{8})z^{6}\nonumber \\
&&-4(13312-6304\delta^{2}-872\delta^{4}-3\delta^{6})z^{7}+8(4512-500\delta^{2}-15\delta^{4})z^{8}\nonumber \\
&& -32(336-\delta^{2})+1280z^{10}]
\nonumber \\
&&
+\delta^{2}(z-2)^{4}[8\delta^{4}(16+2\delta^{2}+3\delta^{4})-32\delta^{4}(8-\delta^{2})z\nonumber\\
&&-2\delta^{2}(320-272\delta^{2}+64\delta^{4}-3\delta^{6})z^{2}+16\delta^{2}(40-54\delta^{2}-5\delta^{4})z^{3}\nonumber \\
&& -(1024-720\delta^{4}-3\delta^{6})z^{4}+8(96-36\delta^{2}-
5\delta^{4})z^{5} \nonumber \\
&&
+8(80+7\delta^{2})z^{6}-512z^{7}]\ln\frac{z\sqrt{4+\delta^{2}-4z}+2\sqrt{(1-z)(z^{2}-\delta^{2})}}
{z\sqrt{4+\delta^{2}-4z}-2\sqrt{(1-z)(z^{2}-\delta^{2})}}\}.
\end{eqnarray}

\begin{eqnarray}
{\rm
S}_{\chi_{c0}}&=&\frac{8\pi}{9s^{3}\delta^{4}z^{5}(z-2)^{8}(z^{2}-\delta^{2})}
\{-4z\sqrt{(1-z)(z^{2}-\delta^{2})(4+\delta^{2}-4z)}\nonumber\\
&&\times[2304\delta^{10}-1152\delta^{8}(26+5\delta^{2})z+192\delta^{6}(640+464\delta^{2}+35\delta^{4})z^{2}\nonumber\\
&& +96\delta^{4}(1152-4816\delta^{2}-1136\delta^{4}-43\delta^{6})z^{3}\nonumber\\
&& +16\delta^{2}(4608-33024\delta^{2}+44752\delta^{4}+4360\delta^{6}+75\delta^{8})z^{4}\nonumber\\
&& -8\delta^{2}(21504-123392\delta^{2}+78448\delta^{4}+2884\delta^{6}-45\delta^{8})z^{5}\nonumber\\
&& -4(12288-156672\delta^{2}+244224\delta^{4}-78128\delta^{6}-512\delta^{8}+21\delta^{10})z^{6}\nonumber\\
&& -2(24576+549888\delta^{2}-356096\delta^{4}+41744\delta^{6}+80\delta^{8}-9\delta^{10})z^{7}\nonumber\\
&& -8(4608-93952\delta^{2}+45728\delta^{4}-1206\delta^{6}+27\delta^{8})z^{8}\nonumber\\
&& +(487424-208384\delta^{2}+119424\delta^{4}+696\delta^{6}-9\delta^{8})z^{9}\nonumber\\
&& -4(155904+4160\delta^{2}+5216\delta^{4}-21\delta^{6})z^{10} \nonumber \\
&&
+16(22976+1480\delta^{2}+85\delta^{4})z^{11}-480(232+11\delta^{2})z^{12}+15104z^{13}]
\nonumber \\
&& +3\delta^{2}(z-2)^{4}[-192\delta^{10}+96\delta^{8}(26+3\delta^{2})z
-64\delta^{6}(160+75\delta^{2}+3\delta^{4})z^{2}\nonumber\\
&& -16\delta^{4}(576-1664\delta^{2}-183\delta^{4}-2\delta^{6})z^{3}\nonumber\\
&& -4\delta^{2}(1536-4608\delta^{2}+4016\delta^{4}+152\delta^{6}+5\delta^{8})z^{4}\nonumber\\
&& +2\delta^{2}(11264-23424\delta^{2}-160\delta^{4}+106\delta^{6}+3\delta^{8})z^{5}\nonumber\\
&& +4(2048-4224\delta^{2}+9952\delta^{4}+248\delta^{6}-27\delta^{8})z^{6}\nonumber\\
&& -(20480-22528\delta^{2}+5312\delta^{4}-368\delta^{6}+3\delta^{8})z^{7}\nonumber\\
&& +4(4096-6496\delta^{2}-600\delta^{4}+17\delta^{6})z^{8}-16(320-472\delta^{2}+7\delta^{4})z^{9}\nonumber \\
&&+32(48+\delta^{2})z^{10}]\ln\frac{z\sqrt{4+\delta^{2}-4z}+2\sqrt{(1-z)(z^{2}-\delta^{2})}}
{z\sqrt{4+\delta^{2}-4z}-2\sqrt{(1-z)(z^{2}-\delta^{2})}}\}.
\end{eqnarray}

\begin{eqnarray}
{\rm
\alpha}_{\chi_{c0}}&=&\frac{8\pi}{9s^{3}\delta^{4}z^{5}(z-2)^{8}(z^{2}-\delta^{2})}
\{4z\sqrt{(1-z)(z^{2}-\delta^{2})(4+\delta^{2}-4z)}\nonumber\\
&& \times[2304\delta^{10}-5760\delta^{8}(6+\delta^{2})z+192\delta^{6}(896+424\delta^{2}+35\delta^{4})z^{2}\nonumber\\
&& +96\delta^{4}(384-4528\delta^{2}-904\delta^{4}-43\delta^{6})z^{3}\nonumber\\
&&
+16\delta^{2}(1536+3840\delta^{2}+23536\delta^{4}+2992\delta^{6}+75\delta^{8})z^{4}\nonumber\\
&&
-8\delta^{2}(52224+8704\delta^{2}+3280\delta^{4}+1924\delta^{6}-45\delta^{8})z^{5}\nonumber\\
&&+4(12288+70656\delta^{2}-51200\delta^{4}-34224\delta^{6}-232\delta^{8}-21\delta^{10})z^{6}\nonumber\\
&&+2(24576+336896\delta^{2}+133888\delta^{4}+53904\delta^{6}+376\delta^{8}+9\delta^{10})z^{7}\nonumber\\
&&+16(2304-62720\delta^{2}-11280\delta^{4}-2191\delta^{6}-30\delta^{8})z^{8}\nonumber\\
&&-(487424-605696\delta^{2}-61312\delta^{4}-7016\delta^{6}-9\delta^{8})z^{9}\nonumber\\
&&+4(155904-40768\delta^{2}-2560\delta^{4}-21\delta^{6})z^{10}\nonumber\\
&&-16(22976-504\delta^{2}+85\delta^{4})z^{11}+480(232+11\delta^{2})z^{12}-15104z^{13}]
\nonumber \\
&&
+3\delta^{2}(z-2)^{4}[192\delta^{10}-288\delta^{8}(10+\delta^{2})z
+64\delta^{6}(224+59\delta^{2}+3\delta^{4})z^{2}\nonumber\\
&&+16\delta^{4}(192-1248\delta^{2}-121\delta^{4}-2\delta^{6})z^{3}\nonumber\\
&&+4\delta^{2}(512-12288\delta^{2}+2384\delta^{4}-56\delta^{6}+5\delta^{8})z^{4}\nonumber\\
&&+2\delta^{2}(3072+35968\delta^{2}-160\delta^{4}+50\delta^{6}-3\delta^{8})z^{5}\nonumber\\
&&-4(2048-4992\delta^{2}+8224\delta^{4}-408\delta^{6}-23\delta^{8})z^{6}\nonumber\\
&&+(12288-51200\delta^{2}-3008\delta^{4}-1456\delta^{6}-3\delta^{8})z^{7}\nonumber\\
&&-4(2048-8992\delta^{2}-1224\delta^{4}-17\delta^{6})z^{8}+16(192-616\delta^{2}-7\delta^{4})z^{9}\nonumber\\
&&+32(16+\delta^{2})z^{10}]\ln\frac{z\sqrt{4+\delta^{2}-4z}+2\sqrt{(1-z)(z^{2}-\delta^{2})}}
{z\sqrt{4+\delta^{2}-4z}-2\sqrt{(1-z)(z^{2}-\delta^{2})}}\}.
\end{eqnarray}

\begin{eqnarray}
{\rm
S}_{\chi_{c1}}&=&\frac{-8\pi}{3s^{3}\delta^{4}z^{5}(z-2)^{8}(z^{2}-\delta^{2})}
\{4z\sqrt{\frac{(1-z)(z^{2}-\delta^{2})}{(4+\delta^{2}-4z)}}[2304\delta^{8}(3+\delta^{2})(4+\delta^{2})\nonumber\\
&&
-1152\delta^{6}(192+208\delta^{2}+62\delta^{4}+5\delta^{6})z\nonumber\\
&&+192\delta^{4}(3072+6400\delta^{2}+3568\delta^{4}+668\delta^{6}+35\delta^{8})z^{2}\nonumber\\
&&-96\delta^{4}(26624+27808\delta^{2}+9992\delta^{4}+1276\delta^{6}+43\delta^{8})z^{3}\nonumber\\
&&+16\delta^{2}(36864+277248\delta^{2}+195296\delta^{4}+50464\delta^{6}+4406\delta^{8}+75\delta^{10})z^{4}\nonumber\\
&&-8\delta^{2}(258048+521984\delta^{2}+302624\delta^{4}+57800\delta^{6}+2672\delta^{8}-45\delta^{10})z^{5}\nonumber\\
&&-4(98304-753664\delta^{2}-564992\delta^{4}-310048\delta^{6}-37736\delta^{8}-172\delta^{10}+21\delta^{12})z^{6}\nonumber\\
&&+2(983040-659456\delta^{2}+84480\delta^{4}-103008\delta^{6}-9000\delta^{8}-220\delta^{10}+9\delta^{12})z^{7}\nonumber\\
&&-(4784128+2330624\delta^{2}+1528576\delta^{4}+120800\delta^{6}+396\delta^{8}+117\delta^{10})z^{8}\nonumber\\
&&+(6914048+3928064\delta^{2}+1137792\delta^{4}+74544\delta^{6}+1900\delta^{8}-9\delta^{10})z^{9}\nonumber\\
&&-2(3100672+1294336\delta^{2}+200672\delta^{4}+8036\delta^{6}-9\delta^{8})z^{10}\nonumber\\
&&+8(443392+116992\delta^{2}+8048\delta^{4}+35\delta^{6})z^{11}\nonumber\\
&&-64(20288+2808\delta^{2}+51\delta^{4})z^{12}+512(544+33\delta^{2})z^{13}-28672z^{14}]\nonumber \\
&&-3\delta^{2}(z-2)^{4}[-192\delta^{8}(3+\delta^{2})+96\delta^{6}(48+28\delta^{2}+3\delta^{4})z\nonumber\\
&&-16\delta^{4}(768+808\delta^{2}+217\delta^{4}+12\delta^{6})z^{2}\nonumber\\
&&+16\delta^{4}(1600+652\delta^{2}+105\delta^{4}+2\delta^{6})z^{3}\nonumber\\
&&+4\delta^{2}(7168-4352\delta^{2}-360\delta^{4}-59\delta^{6}-5\delta^{8})z^{4}\nonumber\\
&&-2\delta^{2}(24576-3968\delta^{2}+1024\delta^{4}-64\delta^{6}-3\delta^{8})z^{5}\nonumber\\
&&+\delta^{2}(17408-7296\delta^{2}+136\delta^{4}-51\delta^{6})z^{6}\nonumber\\
&&-(8192-12800\delta^{2}-8576\delta^{4}-300\delta^{6}+3\delta^{8})z^{7}\nonumber\\
&&+2(8192-6656\delta^{2}-1328\delta^{4}+17\delta^{6})z^{8}-128(80-10\delta^{2}+\delta^{4})z^{9}\nonumber\\
&&+128(24+5\delta^{2})z^{10}]\ln\frac{z\sqrt{4+\delta^{2}-4z}+2\sqrt{(1-z)(z^{2}-\delta^{2})}}
{z\sqrt{4+\delta^{2}-4z}-2\sqrt{(1-z)(z^{2}-\delta^{2})}}\}.
\end{eqnarray}

\begin{eqnarray}
{\rm
\alpha}_{\chi_{c1}}&=&\frac{-8\pi}{3s^{3}\delta^{4}z^{5}(z-2)^{8}(z^{2}-\delta^{2})}
\{-4z\sqrt{\frac{(1-z)(z^{2}-\delta^{2})}{(4+\delta^{2}-4z)}}[2304\delta^{8}(1+\delta^{2})(4+\delta^{2})\nonumber\\
&&-1152\delta^{6}(64+80\delta^{2}+42\delta^{4}+5\delta^{6})z\nonumber\\
&&+192\delta^{4}(1024+2432\delta^{2}+1360\delta^{4}+404\delta^{6}+35\delta^{8})z^{2}\nonumber\\
&&-96\delta^{4}(8192+11872\delta^{2}+3640\delta^{4}+652\delta^{6}+43\delta^{8})z^{3}\nonumber\\
&&+16\delta^{2}(110592+58624\delta^{2}+71328\delta^{4}+12864\delta^{6}+1522\delta^{8}+75\delta^{10})z^{4}\nonumber\\
&&-8\delta^{2}(724992-245504\delta^{2}-21024\delta^{4}-72\delta^{6}+136\delta^{8}-45\delta^{10})z^{5}\nonumber\\
&&+4(98304+1392640\delta^{2}-1900288\delta^{4}-349344\delta^{6}
-20648\delta^{8}-1756\delta^{10}-21\delta^{12})z^{6}\nonumber\\
&&-2(983040-856064\delta^{2}-5078528\delta^{4}-704352\delta^{6}-37736\delta^{8}-724\delta^{10}-9\delta^{12})z^{7}\nonumber\\
&&+(4784128-7352320\delta^{2}-7412992\delta^{4}-760736\delta^{6}-20452\delta^{8}-447\delta^{10})z^{8}\nonumber\\
&&-(6914048-6197248\delta^{2}-3225472\delta^{4}-202576\delta^{6}-4772\delta^{8}-9\delta^{10})z^{9}\nonumber\\
&&+2(3100672-1243136\delta^{2}-376864\delta^{4}-15260\delta^{6}-9\delta^{8})z^{10}\nonumber\\
&&-8(443392-47360\delta^{2}-10128\delta^{4}+35\delta^{6})z^{11}
+64(20288+472\delta^{2}+51\delta^{4})z^{12}\nonumber\\
&&-512(544+33\delta^{2})z^{13}+28672z^{14}]\nonumber \\
&&+3\delta^{2}(z-2)^{4}[-192\delta^{8}(1+\delta^{2})+96\delta^{6}(16+12\delta^{2}+3\delta^{4})z\nonumber\\
&&-16\delta^{4}(256+344\delta^{2}+59\delta^{4}+12\delta^{6})z^{2}
+16\delta^{4}(448+404\delta^{2}-9\delta^{4}+2\delta^{6})z^{3}\nonumber\\
&&+4\delta^{2}(5120+1792\delta^{2}-856\delta^{4}+135\delta^{6}-5\delta^{8})z^{4}\nonumber\\
&&-2\delta^{2}(16384+6016\delta^{2}-1088\delta^{4}+24\delta^{6}-3\delta^{8})z^{5}\nonumber\\
&&+(32768-25600\delta^{2}-7040\delta^{4}-1864\delta^{6}-57\delta^{8})z^{6}\nonumber\\
&&-(57344-81408\delta^{2}-11904\delta^{4}-884\delta^{6}-3\delta^{8})z^{7}\nonumber\\
&&+2(16384-27648\delta^{2}-2384\delta^{4}-17\delta^{6})z^{8}-128(48-118\delta^{2}-\delta^{4})z^{9}\nonumber\\
&&-128(8+5\delta^{2})z^{10}]\ln\frac{z\sqrt{4+\delta^{2}-4z}+2\sqrt{(1-z)(z^{2}-\delta^{2})}}
{z\sqrt{4+\delta^{2}-4z}-2\sqrt{(1-z)(z^{2}-\delta^{2})}}\}.
\end{eqnarray}

\begin{eqnarray}
{\rm
S}_{\chi_{c2}}&=&\frac{-8\pi}{9s^{3}\delta^{4}z^{5}(z-2)^{8}(z^{2}-\delta^{2})}
\{4z\sqrt{\frac{(1-z)(z^{2}-\delta^{2})}{(4+\delta^{2}-4z)}}\nonumber\\
&&\times[2304\delta^{6}(4+\delta^{2})(144+57\delta^{2}+5\delta^{4})
-1152\delta^{6}(6336+3424\delta^{2}+558\delta^{4}+25\delta^{6})z\nonumber\\
&&-192\delta^{4}(12288-82496\delta^{2}-39168\delta^{4}-5180\delta^{6}-175\delta^{8})z^{2}\nonumber\\
&&+96\delta^{4}(125952-175584\delta^{2}-80408\delta^{4}-8852\delta^{6}-215\delta^{8})z^{3}\nonumber\\
&&+16\delta^{2}(73728-1579776\delta^{2}+532640\delta^{4}+310240\delta^{6}+29834\delta^{8}+375\delta^{10})z^{4}\nonumber\\
&&-8\delta^{2}(651264-3396352\delta^{2}+224480\delta^{4}+323960\delta^{6}+24824\delta^{8}-225\delta^{10})z^{5}\nonumber\\
&&-4(98304-2469888\delta^{2}+3741440\delta^{4}-280928\delta^{6}
-327288\delta^{8}\nonumber\\
&&-15148\delta^{10}+105\delta^{12})z^{6}+2(1179648-5492736\delta^{2}\nonumber\\
&&+1172992\delta^{4}-796064\delta^{6}
-273016\delta^{8}-9940\delta^{10}+45\delta^{12})z^{7}\nonumber\\
&&-(7471104-8568832\delta^{2}-2286336\delta^{4}-864288\delta^{6}-131084\delta^{8}-1377\delta^{10})z^{8}\nonumber\\
&&+(14909440-4112384\delta^{2}-1213056\delta^{4}-33264\delta^{6}+1164\delta^{8}-45\delta^{10})z^{9}\nonumber\\
&&-2(9654272+318976\delta^{2}+139168\delta^{4}+39524\delta^{6}+447\delta^{8})z^{10}\nonumber\\
&&+8(1980416+242048\delta^{2}+25120\delta^{4}+883\delta^{6})z^{11}\nonumber\\
&&-64(119296+10832\delta^{2}+245\delta^{4})z^{12}+1024(1840+73\delta^{2})z^{13}-188416z^{14}]\nonumber \\
&&-3\delta^{2}(z-2)^{4}[-192\delta^{6}(144+57\delta^{2}+5\delta^{4})
+96\delta^{6}(1008+304\delta^{2}+15\delta^{4})z\nonumber\\
&&+16\delta^{4}(4224-6392\delta^{2}-1731\delta^{4}-60\delta^{6})z^{2}\nonumber\\
&&-16\delta^{4}(13632-916\delta^{2}-705\delta^{4}-10\delta^{6})z^{3}\nonumber\\
&&-4\delta^{2}(15360-56448\delta^{2}-8648\delta^{4}+433\delta^{6}+25\delta^{8})z^{4}\nonumber\\
&&+2\delta^{2}(96256-29568\delta^{2}-13280\delta^{4}-340\delta^{6}+15\delta^{8})z^{5}\nonumber\\
&&+(16384-193536\delta^{2}-28672\delta^{4}+14680\delta^{6}+399\delta^{8})z^{6}\nonumber\\
&&-5(8192-11776\delta^{2}-3712\delta^{4}+604\delta^{6}+3\delta^{8})z^{7}\nonumber\\
&&+2(10240+9728\delta^{2}-2784\delta^{4}-79\delta^{6})z^{8}+512(4-47\delta^{2}+2\delta^{4})z^{9}\nonumber\\
&&+256(12+19\delta^{2})z^{10}]\ln\frac{z\sqrt{4+\delta^{2}-4z}+2\sqrt{(1-z)(z^{2}-\delta^{2})}}
{z\sqrt{4+\delta^{2}-4z}-2\sqrt{(1-z)(z^{2}-\delta^{2})}}\}.
\end{eqnarray}

\begin{eqnarray}
{\rm
\alpha}_{\chi_{c2}}&=&\frac{8\pi}{9s^{3}\delta^{4}z^{5}(z-2)^{8}(z^{2}-\delta^{2})}
\{4z\sqrt{\frac{(1-z)(z^{2}-\delta^{2})}{(4+\delta^{2}-4z)}}\nonumber\\
&&\times[2304\delta^{6}(4+\delta^{2})(48+3\delta^{2}+5\delta^{4})
-1152\delta^{6}(2112+576\delta^{2}+170\delta^{4}+25\delta^{6})z\nonumber\\
&&-192\delta^{4}(6144-33088\delta^{2}-7296\delta^{4}-1428\delta^{6}-175\delta^{8})z^{2}\nonumber\\
&&+96\delta^{4}(64512-102688\delta^{2}-16232\delta^{4}-1860\delta^{6}-215\delta^{8})z^{3}\nonumber\\
&&-16\delta^{2}(24576+1006848\delta^{2}-639968\delta^{4}-83584\delta^{6}-5006\delta^{8}-375\delta^{10})z^{4}\nonumber\\
&&+8\delta^{2}(552960+3044608\delta^{2}-1287968\delta^{4}-209864\delta^{6}-6608\delta^{8}+225\delta^{10})z^{5}\nonumber\\
&&+4(98304-4091904\delta^{2}-5157120\delta^{4}+2831904\delta^{6}
+398216\delta^{8}\nonumber\\
&&+2884\delta^{10}-105\delta^{12})z^{6}-2(1179648-15978496\delta^{2}\nonumber\\
&&-5100032\delta^{4}+4153696\delta^{6}
+362376\delta^{8}+4228\delta^{10}-45\delta^{12})z^{7}\nonumber\\
&&+(7471104-38830080\delta^{2}-5298944\delta^{4}+3212896\delta^{6}+207588\delta^{8}-285\delta^{10})z^{8}\nonumber\\
&&-(14909440-29458432\delta^{2}-3163776\delta^{4}+769936\delta^{6}+15836\delta^{8}-45\delta^{10})z^{9}\nonumber\\
&&+2(9654272-5982720\delta^{2}-388704\delta^{4}+34020\delta^{6}+447\delta^{8})z^{10}\nonumber\\
&&-8(1980416-193664\delta^{2}-9312\delta^{4}+883\delta^{6})z^{11}\nonumber\\
&&+64(119296+3792\delta^{2}+245\delta^{4})z^{12}-1024(1840+73\delta^{2})z^{13}+188416z^{14}]\nonumber \\
&&-3\delta^{2}(z-2)^{4}[-192\delta^{6}(48+3\delta^{2}+5\delta^{4})
+288\delta^{6}(112+8\delta^{2}+5\delta^{4})z\nonumber\\
&&+16\delta^{4}(1920-3784\delta^{2}+71\delta^{4}-60\delta^{6})z^{2}\nonumber\\
&&-16\delta^{4}(6336-3660\delta^{2}+289\delta^{4}-10\delta^{6})z^{3}\nonumber\\
&&-4\delta^{2}(13312-43392\delta^{2}+9928\delta^{4}-805\delta^{6}+25\delta^{8})z^{4}\nonumber\\
&&+2\delta^{2}(92160-76928\delta^{2}+6944\delta^{4}-676\delta^{6}+15\delta^{8})z^{5}\nonumber\\
&&+(16384-322560\delta^{2}+121856\delta^{4}+10920\delta^{6}+365\delta^{8})z^{6}\nonumber\\
&&-(24576-248320\delta^{2}+96896\delta^{4}+5332\delta^{6}-15\delta^{8})z^{7}\nonumber\\
&&+2(14336-40960\delta^{2}+12768\delta^{4}+79\delta^{6})z^{8}-512(36-55\delta^{2}+2\delta^{4})z^{9}\nonumber\\
&&-256(4+19\delta^{2})z^{10}]\ln\frac{z\sqrt{4+\delta^{2}-4z}+2\sqrt{(1-z)(z^{2}-\delta^{2})}}
{z\sqrt{4+\delta^{2}-4z}-2\sqrt{(1-z)(z^{2}-\delta^{2})}}\}.
\end{eqnarray}

\newpage

\newpage

\newpage

\newpage

\newpage

\newpage

\newpage

\newpage

\newpage

\end{document}